\documentclass[
pra,superscriptaddress,10pt,letterpaper]
{revtex4}

\usepackage{graphicx,amsmath,amsfonts,algorithm}
\usepackage{graphics}
\usepackage{amssymb}
\usepackage{enumitem}
\usepackage{subfigure}

\usepackage[letterpaper,bindingoffset=0in,left=1in,right=1in,top=0.7in,bottom=1.1in,footskip=.425in,headsep=0.125in,headheight=1.5in]{geometry}
\usepackage{mathptmx} 
\usepackage{fancyhdr}
\usepackage{anyfontsize}

\usepackage{siunitx} 

\usepackage[nonumberlist]{glossaries}
\newacronym[shortplural={BC}]{BC}{BC}{beam combiner}
\newacronym[shortplural={BS}]{BS}{BS}{beam splitter}
\newacronym{COST}{COST}{Cooperation in Science and Technology} 
\newacronym{DE}{DE}{detection efficiency}
\newacronym{DES}{DES}{Data Encryption Standard}
\newacronym{DSA}{DSA}{Digital Signature Algorithm}
\newacronym{ECDSA}{ECDSA}{Elliptic Curve Digital Signature Algorithm}
\newacronym{EPSRC}{EPSRC}{Engineering and Physical Sciences Research Council}
\newacronym{InGaAs}{InGaAs}{Indium Gallium Arsenide}
\newacronym{LiNbO3}{LiNbO\ensuremath{_\textrm{3}}}{Lithium Niobate}
\newacronym{MPN}{MPN}{mean photon number}
\newacronym{PBC}{PBC}{polarisation beam combiner}
\newacronym[longplural={polarisation beam splitters},shortplural={PBS}]{PBS}{PBS}{Polarisation Beam Splitter}
\newacronym{PM}{PM}{polarisation-maintaining}
\newacronym{PMF}{PMF}{polarisation-maintaining fibre}
\newacronym{QBER}{QBER}{quantum bit error rate}
\newacronym[longplural={quantum digital signatures},shortplural={QDS}]{QDS}{QDS}{quantum digital signature}
\newacronym{QE}{QE}{quantum efficiency}
\newacronym{QKD}{QKD}{quantum key distribution}
\newacronym{qubit}{qubit}{qubit}
\newacronym{QuTiP}{QuTiP}{Quantum Toolbox in Python}
\newacronym{RSA}{RSA}{Rivest, Shamir, Adleman}
\newacronym{Si}{Si}{Silicon}
\newacronym{Si-SPAD}{Si-SPAD}{silicon single-photon avalanche diode}
\newacronym{SPAD}{SPAD}{single-photon avalanche diode}
\newacronym{SPC}{SPC}{static polarisation controller}
\newacronym{USD}{USD}{unambiguous state discrimination}
\newacronym{USE}{USE}{unambiguous state elimination}
\newacronym{VCSEL}{VCSEL}{vertical cavity surface-emitting laser}

\def\ket#1{| #1 \rangle}

\newtheorem{prop}{Proposition}\def\PRO{\begin{prop}}\def\ORP{\end{prop}}
\newtheorem{coro}{Corollary}\def\COR{\begin{coro}}\def\ROC{\end{coro}}
\newtheorem{theo}{Theorem}\def\TH{\begin{theo}}\def\HT{\end{theo}}
\def\TH{\begin{theo}}\def\HT{\end{theo}}
\newtheorem{defi}[prop]{Definition}\def\DE{\begin{defi}}\def\ED{\end{defi}}
\newtheorem{lemme}[prop]{Lemma}\def\LE{\begin{lemme}}\def\EL{\end{lemme}}

\newcommand{\textsubscript}[1]{\ensuremath{_\textrm{#1}}}  

\newcommand{\djj}{d\kern-0.4em\char"16\kern-0.1em}

\begin{document}
\title{Experimental demonstration of kilometer-range quantum digital signatures}

\author{Ross J. Donaldson}
\affiliation{SUPA, Institute of Photonics and Quantum Sciences, School of Engineering and Physical Sciences, David Brewster Building, Heriot-Watt University, Edinburgh, EH14 4AS, United Kingdom}

\author{Robert J. Collins}
\affiliation{SUPA, Institute of Photonics and Quantum Sciences, School of Engineering and Physical Sciences, David Brewster Building, Heriot-Watt University, Edinburgh, EH14 4AS, United Kingdom}

\author{Klaudia Kleczkowska}
\affiliation{SUPA, Institute of Photonics and Quantum Sciences, School of Engineering and Physical Sciences, David Brewster Building, Heriot-Watt University, Edinburgh, EH14 4AS, United Kingdom}

\author{Ryan Amiri}
\affiliation{SUPA, Institute of Photonics and Quantum Sciences, School of Engineering and Physical Sciences, David Brewster Building, Heriot-Watt University, Edinburgh, EH14 4AS, United Kingdom}

\author{Petros Wallden}
\affiliation{School of Informatics, Informatics Forum, University of Edinburgh, 10 Crichton Street, Edinburgh, EH8 9AB, United Kingdom}

\author{Vedran Dunjko}
\affiliation{Institute for Theoretical Physics, University of Innsbruck, Technikerstrasse 25, A-6020 Innsbruck, Austria}
\affiliation{Division of Molecular Biology, Ruder Bo\u{s}kovi\'{c} Institute, Bijeni\u{c}ka cesta 54, 10002 Zagreb, Croatia.}

\author{John Jeffers}
\affiliation{SUPA, Department of Physics, John Anderson Building, University of Strathclyde, 107 Rottenrow, Glasgow, G4 0NG, United Kingdom}

\author{Erika Andersson}
\affiliation{SUPA, Institute of Photonics and Quantum Sciences, School of Engineering and Physical Sciences, David Brewster Building, Heriot-Watt University, Edinburgh, EH14 4AS, United Kingdom}

\author{Gerald Buller}
\affiliation{SUPA, Institute of Photonics and Quantum Sciences, School of Engineering and Physical Sciences, David Brewster Building, Heriot-Watt University, Edinburgh, EH14 4AS, United Kingdom}

\begin{abstract}
We present an experimental realization of a quantum digital signature protocol which, together with a standard quantum key distribution link, increases transmission distance to kilometre ranges, three orders of magnitude larger than in previous realizations. The bit-rate is also significantly increased compared with previous quantum signature demonstrations. This work illustrates that quantum digital signatures can be realized with optical components similar to those used for quantum key distribution, and could be implemented in existing optical fiber networks.
\end{abstract}

\maketitle


\section{Introduction}

Signature schemes are widely used in electronic communication to guarantee the authenticity and transferability of messages. Transferability means that a signed message is unlikely to be accepted by one recipient and, if forwarded, subsequently rejected by another recipient~\cite{STI2006}. This property distinguishes signature schemes from message authentication schemes, in which there is no transferability requirement. Transferability is closely related to non-repudiation; message repudiation would mean that a sender can successfully deny having sent a message they really did send. The most widespread signature schemes are the public key protocols RSA~\cite{RIV1978}, DSA~\cite{Elgamal1985} and ECDSA~\cite{Johnson2001}, in which security depends on the computational difficulty of factorising large numbers or finding discrete logarithms. Since the security of such schemes is not information-theoretic, but relies on computational assumptions, it can be retrospectively affected by future advances in technology or the discovery of efficient algorithms. In fact, all of the above schemes are known to be insecure against an adversary with a quantum computer~\cite{SHO1997}. 

The security of quantum digital signatures (QDS)~\cite{Gottesman2002,Andersson2006}, on the other hand, is information-theoretic, guaranteed by the laws of quantum mechanics to be secure against an adversary with infinite computational capabilities. This is a potentially significant advantage. There also exist unconditionally secure ``classical" digital signature schemes~\cite{Swanson2011}. In addition to requiring pairwise secret classical communication channels (which could be achieved using QKD), the scheme by Chaum and Roijakkers~\cite{Chaum:1990:USD:646755.705359} requires an authenticated broadcast channel (a significantly more challenging prospect), and the one by Hanaoka \textit{et al.}~\cite{Hanaoka2000} is phrased in terms of a third party, trusted by all participants. Such assumptions are not straightforward and are not required in QDS protocols. Pairwise message authentication can be efficiently implemented with information-theoretic security using short pre-shared keys \cite{WEG1981} and is \emph{not} equivalent with the stronger assumption of an authenticated broadcast channel. Further, information-theoretically secure secret classical channels can be generated using QKD. Therefore, even in terms of required resources, if the secret shared keys are generated using QKD, QDS schemes are no more demanding than the ``classical'' unconditionally secure signature schemes.

Although QDS has been successfully realized in the lab between three parties, a sender Alice and two receivers Bob and Charlie~\cite{Clarke2012,Collins2014}, the transmission distance for these realizations was limited to the order of meters. The earliest versions of QDS protocols~\cite{Gottesman2002,Clarke2012} also required long-term quantum memory~\cite{Bussieres2013,Simon2010,Specht2011,Reim2011} to store the signature states at the receivers, making full implementation difficult with currently available technology. The requirement for quantum memory is removed if the recipients directly measure the quantum states sent by Alice, for example using unambiguous state elimination (USE)~\cite{Dunjko2014,Wallden2013}, and then storing only the classical measurement outcomes~\cite{Collins2014,Dunjko2014}. However, these QDS schemes still relied on a multiport to guarantee non-repudiation, comprising two intertwined interferometers controlled locally by Bob and Charlie. The multiport design required internal delays equal to the link length between Bob and Charlie, as well as introducing unavoidable additional high optical loss, restricting the practical transmission distance to approximately 5~m~\cite{Clarke2012,Collins2014}. 

\section{Description of method}

For QDS to be useful in real world applications, protocols which allow for higher transmission rate and greater distance between parties must be developed and demonstrated experimentally. In this paper we present an experimental realization of a key part of such a protocol, along the lines of new QDS protocols in~\cite{Wallden2015}. The experimental system, outlined in Figure \ref{Fig:experiment} and described in greater detail in the Appendix, uses similar optical components to those currently employed in phase-encoded QKD experiments~\cite{GIS2002,Sasaki2011,Jouguet2013,Singh2014,Clarke2011b}, thus greatly improving the practicality of QDS. 

Signature protocols have two stages, a distribution stage and a messaging stage. The scheme is set up in the distribution stage, to enable signed messages to be sent and received in the messaging stage, which could occur at any future date. An outline of the protocol will be presented here with a more complete description in the Appendix. We will describe how to sign a 1-bit message $m$; longer messages may be sent by suitably iterating this procedure. In the distribution stage, Alice chooses two random sequences of $L$ phase-encoded coherent states, one sequence for each possible message, 0 or 1. Increasing the length $L$ of the sequence will increase the security of the scheme. The security also depends on other parameters, such as the mean photon number per pulse $|\alpha|^2$ and on imperfections in the setup~\cite{Wallden2015}. Alice chooses her quantum states randomly from a known alphabet of non-orthogonal quantum states, in our case the four phase-encoded states $\ket{\alpha}$, $\ket{\alpha e^{i \pi / 2}}$, $\ket{\alpha e^{i \pi }}$, and $\ket{\alpha e^{3i \pi / 2}}$, relative to a common phase reference. She keeps the complete classical description of the two sequences secret; this constitutes her ``private key''. Non-orthogonal quantum states cannot be perfectly distinguished from each other, so only Alice can know her full private key. The phase reference pulse in our implementation is strong, so that tampering with it could in principle be detected by performing state tomography.

Alice sends one copy of each sequence of coherent quantum states to both Bob and Charlie, through separate quantum channels. Bob and Charlie measure the received coherent states, in our case using quantum USE~\cite{Collins2014,Wallden2013,Wallden2015}, ruling out zero, one or more of the four possible phases for each position in each sequence of states. Bob and Charlie perform the measurements by combining the suitably adjusted reference pulses with the signal pulses using two beam splitters, one for each non-orthogonal phase pair in the four-state alphabet. Each detection event eliminates one of the four possible states sent by Alice. In Figure \ref{Fig:experiment}, the phase of the reference state entering beam splitter 2 at Bob is set so that he can eliminate the $0$ and $\pi$ phases, and the phase at beam splitter 3 is set to eliminate the $\pi / 2$ and $3\pi / 2$ phases. From the detection statistics, one can calculate the conditional probabilities for Bob to eliminate each of the four states, given that Alice sent a particular state. An example of this so-called cost matrix is illustrated in Figure \ref{Fig:mainResults}(a) for $|\alpha|^2 = 0.5$.

Bob and Charlie now each have a measurement record for the sequences sent by Alice. They then both randomly and independently choose half of their measurement outcomes to forward to the other recipient. They keep secret from Alice which measurement outcomes are forwarded and which are kept. This last step is not implemented in our present setup, but could be achieved using a standard QKD link. Employing a standard QKD link does not have an impact on the maximum transmission distance or security level of the system and does not add further complications in the security analysis. This is because it is only used to secure the classical processing after the coherent states have been transmitted and measured. The random forwarding procedure replaces the symmetrising multiport in earlier implementations and ensures that Bob and Charlie obtain the same measurement statistics irrespective of what states Alice sends them. This is true even for the most general cheating strategies by Alice, which could involve entangled states. A dishonest Alice could attempt repudiation, that is, deny having sent a message that she actually did send. If the whole signature scheme is to have information-theoretic security, then a secret classical communication channel with information-theoretic security is needed for forwarding the measurement results, since otherwise security against repudiation would not be information-theoretic. At the end of the distribution stage, Bob and Charlie should each be left with different and incomplete descriptions of the signature sent by Alice, which are kept until the messaging stage.

In the messaging stage, Alice chooses a message $m$ and sends it together with her corresponding private key, that is, the classical description of the corresponding sequence of quantum states, to the intended recipient, say Bob. All communication during the messaging stage takes place over pairwise authenticated classical communication channels; quantum communication is needed only in the distribution stage. To accept the message, Bob checks Alice's private key against his measurement record for the message $m$. He accepts the message if he finds fewer than $L s_a$ mismatches, where $s_a$ is an authentication threshold and $L$ is the length of the sequences chosen by Alice.

If Bob wishes to forward the message, he sends the message together with Alice's private key to Charlie. Charlie then checks for mismatches in the same way as before, but applies a different verification threshold $s_v$, which is larger than $s_a$. The message is only accepted if there are fewer than $L s_v$ mismatches. It is important that the threshold for accepting a message directly from Alice is different from the threshold for accepting a forwarded message. Otherwise Alice could repudiate with high probability~\cite{Gottesman2002}. Signing a message uses up the distributed signatures, which cannot be reused. 

Here, as for all existing QDS protocols, we assume that none of the participants are tampering with or eavesdropping on the quantum channels between other participants. It is expected this assumption could be removed by using a parameter estimation procedure analogous to that used in QKD~\cite{Amiri2015}. By declaring (sacrificing) some of the states sent during the distribution stage, participants should be able to estimate the level of eavesdropping, aborting if it is too high. When considering security against forging by either Bob or Charlie, one assumes that the other recipient and Alice are honest, since no protocol can guarantee security if two of the three parties are dishonest and collude. Similarly, when considering security against repudiation by Alice, one assumes that Bob and Charlie are honest. 

\begin{figure}[tbp]
\includegraphics[width=12.9cm]{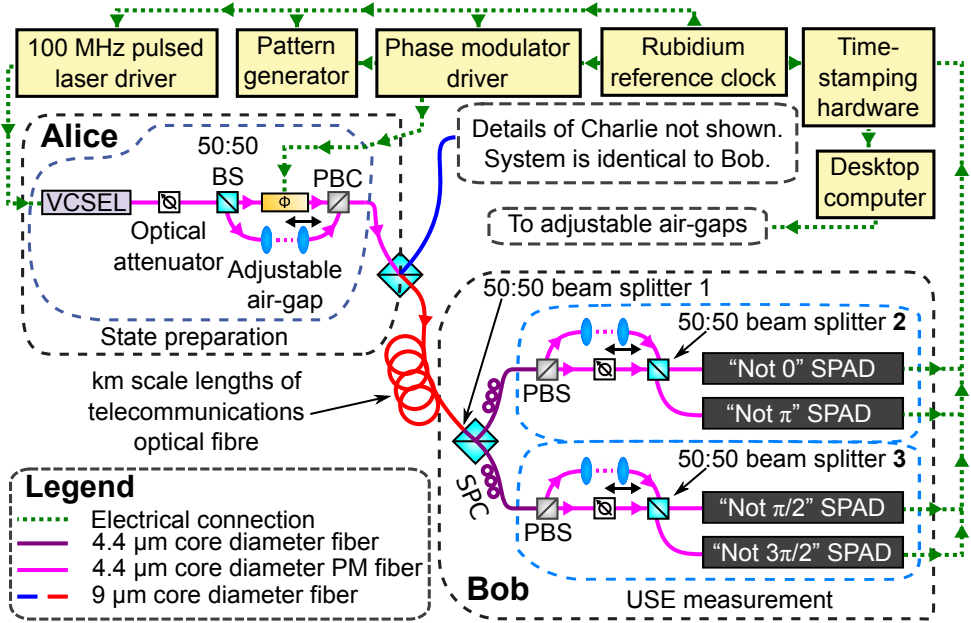}
\caption{(Color online). Experimental setup for kilometer-range quantum digital signatures. Alice uses a pulsed 850~nm wavelength vertical cavity surface emitting laser (VCSEL) diode to generate coherent states. She encodes her phase shift by using a lithium niobate (LiNbO\textsubscript{3}) phase modulator, choosing randomly between the four possible phases. Alice sends the coherent states through standard telecommunications optical fiber to the receivers, who then measure the received coherent states. Sender and receivers are constructed from 4.4~$\mu$m core diameter polarization-maintaining optical fiber to improve the interferometric visibility. Polarization routing of modulated signal and unmodulated reference was carried out using polarization dependent beam splitters (PBS), and static polarization controllers (SPCs) were used to correct for polarization shift induced by birefringence in the telecommunications fiber. \label{Fig:experiment}}
\end{figure}

All pairwise classical communication in the present protocol, just as for QKD, must be authenticated. Pairwise message authentication can in modern cryptography be efficiently implemented using ``short'' pre-shared keys~\cite{CAR1979}. It is not, even in principle, possible to prevent man-in-the-middle-attacks in QKD or QDS schemes unless there has been some prior interaction between parties. If information-theoretic security is required, one needs to use an appropriate authentication scheme for all classical communication~\cite{Abidin2014}. The security analysis for the QDS protocol implemented here proceeds much as in Collins \textit{et al.}~\cite{Collins2014} and is detailed in the Appendix.

\section{Experimental results}

In Figure \ref{Fig:mainResults}, we plot experimental data and parameters for a range of mean photon numbers $|\alpha|^2$ sent by Alice. When $|\alpha|^2$ increases, the probability that a forger will be able to select the correct declaration increases, but likewise does an honest recipient's ability to detect mismatches in a fake declaration. It is therefore highly non-trivial to determine the optimal value of $|\alpha|^2$. The gap, $g$, between the probability for a forger's fake declaration to be rejected and the probability for Alice's true declaration to be rejected, i.e. $g = C_{min}  - p_h$ (see the Appendix), depends on $|\alpha|^2$ and we choose the optimal value of $|\alpha|^2$ that maximises $g$. Other experimental parameters such as system loss and interferometric visibility are also taken into account in the cost matrices that are used to obtain $C_{min}$ and thus the gap $g$. It can be seen from Figure \ref{Fig:mainResults}(c) that maximum gap occurs around $|\alpha|^2 = 0.4$ for transmission distances of 500 m, 1000 m and 2000 m, showing that this value of $|\alpha|^2$ is the best choice for the sender in this particular experimental implementation. 

For the desired security level, the length required to sign one half-bit can then be calculated using $P(\textrm{protocol failure})=2e^{-\left( g / 4 \right)^2 L}$ (see the Appendix). In this paper, we have used a security level of 0.01\%. The resulting signature length for each half-bit is graphed in Figure \ref{Fig:mainResults}(d) for varying $|\alpha|^2$. As $|\alpha|^2$ increases, the length per half-bit dips to a minimum and increases again. For our experimental data, the maximum value for the gap and the minimum values for the length are around $|\alpha|^2 = 0.4$. 

\begin{figure}[tbp]
\includegraphics[width=12.9cm]{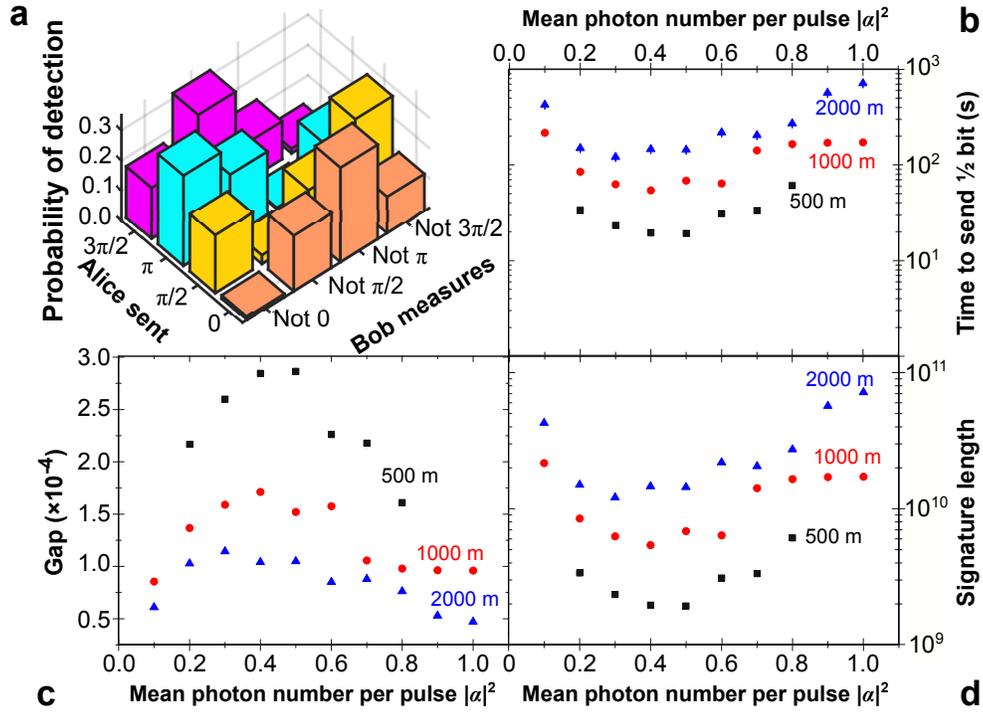}
\caption{(Color online). Experimental results and calculated parameters for one of the recipients, Bob. In our experiment, measurement statistics for Bob and Charlie was very similar. a) Conditional probabilities for unambiguous state elimination by Bob, with $|\alpha|^2 = 0.5$. These probabilities are used to calculate the gap $g$ and the required signature length per half-bit. b) the time required for Alice to send a half-bit to Bob. c) The gap achieved in the experiments is a calculated from a combination of information from both receivers' cost matrices. d) The length $L$ required to prevent forging and repudiation, per half-bit of message, for a security level of 0.01\%. For all sub-figures at 500~m, mean photon number points of 0.1, 0.9 and 1 have been removed. Here, the combination of count rate and the dead time of the detector led to non-linearity in the detection.\label{Fig:mainResults}}
\end{figure}

\begin{figure}[tbp]
\includegraphics[width=12cm,trim=0.20cm 0.125cm 0.5cm 0.5cm, clip=true]{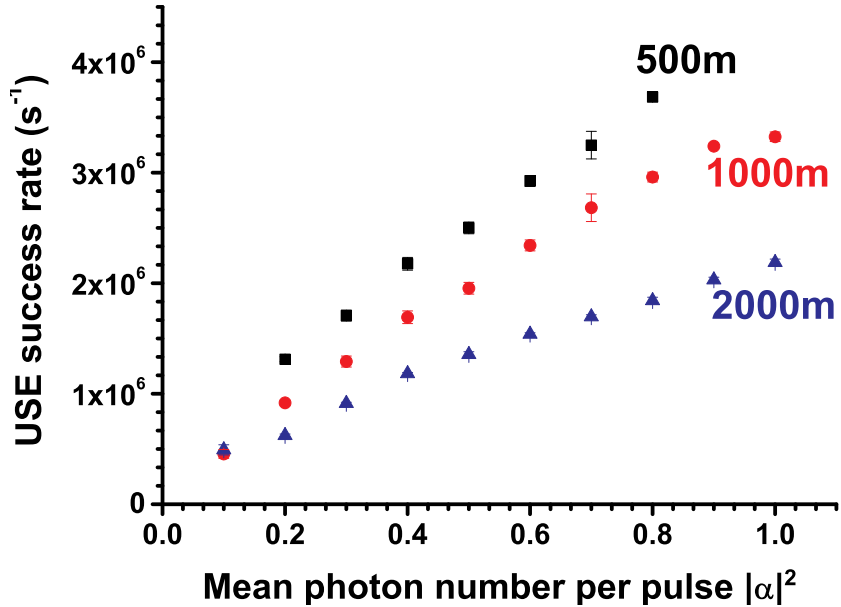}
\caption{(Color online). Unambiguous state elimination (USE) success rate at different transmission distances for Bob. Results for Bob and Charlie are again very similar so only results for Bob are shown. The failure rate of USE is on average only 1.7\% of the pulse-repetition frequency and is not plotted. For transmission distance 500~m, points for mean photon number 0.1, 0.9 and 1 have been removed, since here, the combination of count rate and the dead time of the detector led to non-linearity in the detection of events. \label{Fig:successResults}}
\end{figure}

To further illustrate the measurements by the recipients, experimental success rates are shown in Figure \ref{Fig:successResults}. A successful USE measurement means eliminating any state except the state Alice sent. On average, the USE success rate is 80\% of the raw count rate. Only a single detection event is required to eliminate a state, and therefore the success rate is higher than for unambiguous state discrimination (USD), where one has to eliminate all possible states except the one sent. The success rate for USD is typically many orders of magnitude lower than the time-gated detector count rate~\cite{Collins2014}. The failure rate of USE (eliminating the state that was actually sent) depends on the visibility of the interferometers formed by Alice's state preparation and the receiver's detection setup. In the security proof we assume that Alice has full control over the failure rate and so can generate any number of mismatches she likes. For these experiments, the USE failure rate was on average 1.7\% of the pulse repetition rate. Knowing Alice's repetition rate and the signature length $L$ required to sign a half-bit message allows us to calculate the time it would take for Alice to send one half-bit using the current system. 

\section{Discussion}

In a previous QDS experiment using quantum state elimination and a mulitport~\cite{Collins2014}, with a transmission distance of 5 m, the largest gap was found to be $1.20 \times 10^{-6}$, which occurred for $|\alpha|^2 = 1$. The USE success rate was $2 \times 10^5$ counts per second. For a security level of 0.01\%, the length $L$ required to securely sign a half-bit was found to be $5.0 \times 10^{13}$, and the estimated time it would take to securely sign a half-bit was over eight years. In contrast, for the present setup and a transmission distance of 500~m, the maximum gap, $g = 2.86 \times 10^{-4}$, occurs at a $|\alpha|^2 = 0.5$, with a USE success rate of $2.48 \times 10^6$ . This gap is two orders of magnitude greater than for the previous short-range system. The length $L$ per half-bit for this new realization is $1.93 \times 10^{9}$ for a security level of 0.01\%, and the estimated time it would take to securely sign a half-bit is less than 20~seconds. As the transmission distance increases, the time taken to sign one half-bit naturally increases, but even at 2000~m the time taken to sign one half-bit is four orders of magnitude smaller than in the previous experimental demonstrations of QDS.

The system reported in this letter operates over relatively short distances when compared to the current maximum transmission distances achieved for QKD experiments~\cite{Korzh2014}\.  Quite apart from the increased losses of 2.2~dbkm\textsuperscript{-1} experienced in the standard telecommunications fibre quantum channel at the wavelength of 850~nm selected for these experiments as opposed to 0.2~dbkm\textsuperscript{-1} at the wavelength of 1550~nm typically used in the QKD experiments, this letter reports a QDS protocol and it is to previous experimental QDS systems that it must be compared. In particular, because of differences in the security analysis, one would not expect, a-priori, that the exact same physical implementation leads to comparable distances of QKD and QDS protocols.

Although the new protocol demonstrated in this paper shows a significant enhancement in transmission distance and message signing rate compared with previous QDS demonstrations, further improvements in signature transmission rate, in particular, will be necessary to fully exploit the potential of QDS. The largest scope for improvement seems to lie in having Alice send different sequences of states to Bob and Charlie. This reduces each participant's ability to forge, while retaining security against repudiation due to the exchange of measurement outcomes performed by Bob and Charlie~\cite{Amiri2015}.

The recently proposed signature protocol ``P2" in~\cite{Wallden2015} uses standard QKD systems to first generate pairwise shared keys between all parties. The secret keys are then used to sign messages. Currently, such schemes seem more efficient than any proposed quantum protocol that directly signs a message without first distilling a secret key. Nevertheless, ``direct" quantum signature protocols, such as the one discussed in this paper, are still worth investigating, as they may eventually prove better than signature protocols based on secret shared keys. For example, direct quantum protocols can remain secure even if the available quantum channels are too noisy for practical QKD~\cite{Amiri2015}. The most efficient direct quantum protocol should intuitively be at least as efficient as protocols proceeding via first distilling shared secret keys, and then using the shared keys to sign messages. There could also be other advantages with direct quantum protocols. If quantum channels of sufficiently high quality are available, then direct quantum signature protocols should need quantum channels only between the sender and each recipient, as opposed to QKD links between each party for protocols similar to P2. This means that the number of quantum channels scales only linearly with the number of recipients for direct quantum signature schemes, as opposed to quadratically for schemes similar to P2. Compared with work on QKD, very little work has been done on QDS schemes. This is surprising, given the wide usage and importance of signature schemes in modern communication. We are confident that demonstrating kilometer-range quantum signature schemes, using the same technical components as QKD, will stimulate wider interest in developing the next generation of quantum signature protocols for real-world optical networks.

\acknowledgments
This work was supported by the UK Engineering and Physical Sciences Research Council (EPSRC) under EP/G009821/1, EP/K022717/1, and EP/K015338/1. P.W. gratefully acknowledges support from the European  European Cooperation in Science and Technology (COST) Action MP1006. V.D. acknowledges support by EPSRC.  

\pagebreak
\setcounter{section}{0}
\begin{center}
\vspace{12 pt}
{\Large \textbf{APPENDIX}}
\end{center}
\section{Protocol} \label{sec:protocol}

The protocol presented in this paper is similar to P1' from reference \cite{Wallden2015}, but with two important differences. First, we require Bob and Charlie to record whether each measurement outcome they hold came directly from Alice or whether it was forwarded to them by the other recipient. Second, the experiment uses weak coherent states instead of true single-photon states. This means that the security proofs presented for protocol P1' do not apply, and instead we follow the methods applied in \cite{Collins2014} for security against individual and collective forging attacks. The protocol is as follows:

\subsection{Distribution Stage}
\begin{enumerate}
\item For each future one-bit message $k = 0, 1$, Alice generates two copies of sequences of coherent states, $QuantSig_k = \bigotimes^L_{l=1} \rho^k_l$ where $\rho^k_l = |b^k_l\alpha\rangle\langle b^k_l\alpha|$, $\alpha$ is a real positive amplitude, $b^k_l \in \{1, i, -1, -i\}$ are randomly chosen and $L$ is a suitably chosen integer (the signature length). The state $QuantSig_k$ and the sequence of numbers $PrivKey_k = (b^k_1, ..., b^k_L)$ are called the \emph{quantum signature} and the \emph{private key}, respectively, for message $k$.
\item Alice sends one copy of $QuantSig_k$ to Bob and one to Charlie, for each possible message $k = 0$ and $k = 1$.
\item Bob and Charlie measure each state received using quantum unambiguous state elimination (USE) for $\{ |\alpha\rangle, |i\alpha\rangle, |-\alpha\rangle, |-i\alpha\rangle \}$. For every element of each quantum signature (for $k=0,1$), they store which detectors detected photons; each detector rules out one possible phase state. They therefore store sets of six numbers (hexaplets) of the form $\{k, l^\prime, a^{k, l^\prime}_0, a^{k, l^\prime}_{\pi/2}, a^{k, l^\prime}_\pi, a^{k, l^\prime}_{3\pi/2} \}$, where $1\leq l^\prime \leq L$ and $a^{k, l^\prime}_\phi \in \{0, 1\}$. Here, $a^{k, l^\prime}_\phi = 0$ means that no photons were detected at the $\neg\phi$ detector (that is, the phase $\phi$ is not ruled out), while $a^{k, l^\prime}_\phi=1$ means that there was at least one photon detected at the $\neg\phi$ detector (that is, the phase $\phi$ is ruled out for this element). By $\neg\phi$ detector we symbolise the ``not $\phi$" detector. Note that, due to losses, many of the hexaplets will have $a^{k, l^\prime}_\phi = 0$ for all $\phi$.
\item Once all states have been received and all hexaplets recorded, Bob and Charlie each independently choose half ($L/2$) of their hexaplets and send them to the other participant. To do this, they use a secret classical channel so that Alice has no information regarding which participant holds which hexaplet. Bob and Charlie keep a record of whether a particular hexaplet came from Alice or whether it was forwarded by the other recipient.
\end{enumerate}

\subsection{Messaging Stage}
\begin{enumerate}
\item To send a signed one-bit message $m$, Alice sends $(m, PrivKey_m)$ to the desired recipient (say Bob).
\item Bob checks whether $(m, PrivKey_m)$ matches both of his stored sequences -- the one he received directly from Alice, and the one he received from Charlie. In particular, he counts the number of elements of $PrivKey_m$ which disagree with his stored hexaplets. Therefore, for a given element $l$ of the signature, if Alice's declaration was $\phi$, Bob needs to check if $a_\phi$ is 0 or 1. If $a_\phi=1$, he registers one mismatch. In other words, a mismatch is registered whenever Alice's declaration for a given element has been eliminated by Bob's USE measurement. Bob checks, for each of his stored sequences, whether the number of mismatches is below $s_aL/2$, where $s_a$ is an \textbf{authentication threshold}. Bob accepts the message only if both of his sequences have fewer than $s_aL/2$ mismatches, i.e. both the sequence received directly from Alice, and the sequence received from Charlie.
\item To forward the message to Charlie, Bob forwards to Charlie the pair $(m, PrivKey_m)$ he received from Alice. Charlie tests for mismatches similarly to Bob, but to protect against repudiation by Alice, he uses a different threshold. For each of his sequences -- the one received directly from Alice and the one received from Bob -- Charlie checks if the number of mismatches is below $s_vL/2$, where $s_v$ is the \textbf{verification threshold}, with $s_a < s_v$. Charlie accepts the message only if both of his sequences have fewer than $s_vL/2$ mismatches.
\end{enumerate}

\section{Security} \label{sec:security}

The quantum signature protocol is designed to be secure against two types of dishonesty: forging and repudiation. Security against forging requires that any participant receiving a message will, with high probability, reject a message that did not originate with Alice. Security against repudiation requires that, with high probability, Alice cannot make Bob and Charlie disagree on the validity of her message, i.e. she cannot make one participant accept and the other reject her message. In this section we will show that the probabilities of forging and repudiation decay exponentially in the signature length, $L$. Further, we will show that the protocol is robust, i.e. if all participants are honest, the protocol works and does not abort, except with a probability exponentially small in $L$.

For forging one can distinguish different types of malicious attacks. In individual attacks, the cheating party employs a strategy separately and independently for each signature element. In collective attacks, there may be classical correlations between strategies for different signature elements. Coherent attacks are the most general type; here a cheating party can employ any type of correlations, including entanglement and measurements in an entangled basis. Note that the same distinctions apply in principle to repudiation attempts from Alice. However, when proving security against repudiation we will assume that Alice can exactly control the number of mismatches her signature generates with Bob's and Charlie's measurement outcomes. This covers all types of attack from Alice and so the distinction between individual, collective and coherent attacks is not made. Security is proven for all types of repudiation attacks, and all types of forging except coherent forging attacks. We will treat individual forging attacks in detail. Security against collective forging attacks then follows because the optimal collective forging strategy is actually an individual strategy, as argued in \cite{Collins2014}.

\subsection{Model Assumptions}

Here, as in all previous QDS protocols, we assume that the quantum channels between participants do not allow tampering or eavesdropping from a third party. It is expected that this assumption could be removed by using a parameter estimation procedure analogous to that used in QKD \cite{Amiri2015}. By sacrificing part of the states sent during the distribution stage, participants should be able to estimate the level of tampering and eavesdropping. 

All pairwise classical communication in the present protocol, just as for QKD, must be authenticated. Pairwise message authentication can be efficiently implemented with information-theoretic security using short pre-shared keys \cite{WEG1981}. Without this authentication, it is not possible, even in principle, to prevent ``man-in-the-middle" attacks in QKD or QDS schemes . Thus neither QKD nor QDS can be implemented without some prior interaction between parties. It is important to note that pairwise authentication between parties is far simpler to achieve than an authenticated broadcast channel.

\subsection{Security Against Repudiation}

To repudiate, Alice aims to send a declaration $(m, Sig_m)$ which Bob will accept and which Charlie will reject. To do this, Bob must accept both the elements that Alice sent directly to him, and the elements that Charlie forwarded to him. In order for Charlie to reject he need only reject either the elements he received from Alice, or the elements Bob forwarded to him. Intuitively, security against repudiation follows because of the symmetrisation of measurement outcomes performed by Bob and Charlie using the secret classical channel. Bob and Charlie perform USE measurements on each of the states sent to them by Alice so that they hold the $L$ hexaplets $(b_1, ..., b_L)$ and $(c_1, ..., c_L)$ respectively. We give Alice full powers and assume that later on, in the messaging stage, she is able to fully control the number of mismatches her signature declaration, $PrivKey_m$, contains with the hexaplets $(b_1, ..., b_L)$ and $(c_1, ..., c_L)$. Call the mismatch rates $e_B$ and $e_C$ respectively. The symmetrisation process means that Bob and Charlie will randomly (and unknown to Alice) receive $L/2$ of the others' hexaplets. We show that all choices of $e_B$ and $e_C$ lead to an exponentially decaying probability of repudiation.

Suppose Alice chooses $e_C > s_a$. In this case, Bob is selecting (without replacement) $L$/2 elements from the set $\{c_1, ..., c_L\}$, which contains exactly $e_CL$ mismatches with Alice's declaration. The number of mismatches Bob selects then follows a hypergeometric distribution $H(L, e_CL, L/2)$ with expected value $e_CL/2$. For the message to be accepted, Bob must select fewer than $s_aL/2$ errors. The tails of a hypergeometric distribution can be bound, using \cite{Chvatal1979}, to give an inequality with the same form as a Hoeffding Inequality. We bound the probability that Bob selects fewer than $s_aL/2$ mismatches as
\begin{equation*} 
\mathbb{P} (\text{Bob receives fewer than $s_aL/2$} \text{mismatches from Charlie})\leq \exp(-(e_C-s_a)^2L).
\end{equation*}
To repudiate, Alice must make Bob accept the message, which means that Bob must accept both the part received from Alice and the part received from Charlie. Since $\mathbb{P}(A \cap B) \leq \min\{\mathbb{P}(A), \mathbb{P}(B)\}$, the probability of repudiation must be less than or equal to the above expression, and so must also decrease exponentially.

Now suppose $e_C \leq s_a$. In this case, if $e_B>s_a$ the above argument shows that it is highly likely that Bob will reject the message, so we consider only the case where $e_B \leq s_a$. Consider first the set $\{b_1, ..., b_L\}$. We can use the same arguments as above to bound the probability of selecting more than $s_vL/2$ mismatches as
\begin{equation*}
\mathbb{P}(\text{Charlie selects more than $s_vL/2$} \text{mismatches from Bob}) \leq \exp(-(s_v-e_B)^2L).
\end{equation*}
For Alice to successfully repudiate, Charlie must select more than $s_vL/2$ mismatches from either the set $\{b_1, ..., b_L\}$ or the set $\{c_1, ..., c_L\}$. Using $\mathbb{P}(A \cup B) \leq \mathbb{P}(A) + \mathbb{P}(B)$, we can see that, for the choice of $e_B, e_C \leq s_a$, we have
\begin{equation*}
\mathbb{P}(\text{Charlie selects more than $s_vL/2$ mismatches})\leq 2\exp(-(s_v-s_a)^2L).
\end{equation*}
So again, the probability of Alice successfully repudiating decreases exponentially in the size of the signature. Similar to \cite{Wallden2015}, Alice's best strategy would be to pick $e_B = e_C = \frac{1}{2}(s_v+s_a)$, in which case
\begin{equation} \label{eq:rep}
\mathbb{P}(\text{Repudiation}) \leq 2\exp\left(-\frac{1}{4}(s_v-s_a)^2L\right).
\end{equation}

\subsection{Security Against Individual and Collective Forging}

In order to forge, Bob must make a declaration with fewer than $s_vL/2$ errors with the $L/2$ elements Charlie received directly from Alice. To bound the probability of Bob being able to make such a declaration, we will follow the cost matrix analysis performed in \cite{Collins2014}. For a given individual signature element, we define the cost matrix for Bob as a matrix where the rows correspond to which state Alice sent ($|\exp(i\theta)\alpha\rangle$), while the columns correspond to the detectors $D(\neg\theta)$. Each matrix element $C_{i,j}$ can be taken equal to the probability that if the $i$'th state is sent, then Charlie's $j$'th detector clicks. This is because Bob should avoid declaring a phase that Charlie has eliminated. His cost for making a particular declaration will therefore be proportional to the probability that Charlie has ruled out this state. When $|\alpha|^2 = 0.4$, and over a distance of \SI{500}{\metre}, the experiment gives us the cost matrix
\begin{equation} \label{eq:costmat}
C = \begin{pmatrix}
8.41\times 10^{-5} & 1.48\times 10^{-3} & 2.00\times 10^{-3} & 1.11\times 10^{-3} \\
1.15\times 10^{-3} & 6.31\times 10^{-5} & 1.22\times 10^{-3} & 2.84\times 10^{-3} \\
2.41\times 10^{-3} & 1.90\times 10^{-3} & 1.53\times 10^{-4} & 1.30\times 10^{-3} \\
1.15\times 10^{-3} & 2.77\times 10^{-3} & 1.15\times 10^{-3} & 4.49\times 10^{-5} 
\end{pmatrix}.
\end{equation}
From this matrix, we aim to find two important quantities. First, we want to find the honest cost, $p_h$, for Bob. This is the rate at which Charlie would find mismatches even if Bob acts honestly. This situation occurs when Charlie erroneously rules out the state that Alice actually sent. The experimentally found probability of this happening is given by the diagonal elements of $C$, and so the honest cost is simply the average of the diagonal elements, giving $p_h = 8.61\times 10^{-5}$. This rate is important for two reasons: first, the parameter $s_a$ must be set higher than $p_h$ for the protocol to be robust (to avoid aborting due to noise in the honest run); second, to be secure against forging $p_h$ must be smaller than the rate at which Charlie finds mismatches if Bob is dishonest, i.e. being dishonest must carry some positive cost for Bob.

We now consider the case of a dishonest Bob who tries to guess Alice's signature so that he can forge a message to Charlie. We consider only individual and collective forging attacks, where Bob acts on each quantum state individually, but the outcome of his measurement on one quantum state can affect his choice of measurement on the others. Security against coherent forging attempts remains an open question. We want to find $C_{min}$, the minimum possible rate that Bob will declare single signature elements which have been eliminated by Charlie. To do this, we use the cost matrix $C$ and note that $C_{min}$ is the minimum cost associated with the matrix $C$. Finding minimum costs involves finding optimal measurements, which is in general a difficult problem. However, we are able to bound $C_{min}$ following the method in \cite{Collins2014}. To do this, we find the matrices $C^h$, $C^\prime$ and $C^l$, with
\begin{equation*}
C^h = \begin{pmatrix}
8.41\times 10^{-5} & 8.41\times 10^{-5} & 8.41\times 10^{-5} & 8.41\times 10^{-5} \\
6.31\times 10^{-5} & 6.31\times 10^{-5} & 6.31\times 10^{-5} & 6.31\times 10^{-5} \\
1.53\times 10^{-4} & 1.53\times 10^{-4} & 1.53\times 10^{-4} & 1.53\times 10^{-4} \\
4.49\times 10^{-5} & 4.49\times 10^{-5} & 4.49\times 10^{-5} & 4.49\times 10^{-5} 
\end{pmatrix}.
\end{equation*}
This is a constant-row matrix made up of the diagonal elements of the matrix $C$. The minimum cost achievable for this matrix is the honest cost, $p_h$. The matrix
\begin{equation*}
C^\prime =\begin{pmatrix}
0 & 1.40\times 10^{-3} & 1.92\times 10^{-3} & 1.02\times 10^{-3} \\
1.09\times 10^{-3} & 0 & 1.15\times 10^{-3} & 2.78\times 10^{-3} \\
2.26\times 10^{-3} & 1.75\times 10^{-3} & 0 & 1.15\times 10^{-3} \\
1.11\times 10^{-3} & 2.72\times 10^{-3} & 1.11\times 10^{-3} & 0
\end{pmatrix}
\end{equation*}
 is the difference between $C$ and $C^h$. It represents the difference between an honest strategy and a dishonest strategy. The minimum cost for this matrix is the minimum additional cost suffered from acting dishonestly. It is difficult to find the minimum cost for this matrix, so instead we can bound it below by reducing all non-zero elements to the the smallest non-zero element, to get the matrix
\begin{equation*}
C^l =\begin{pmatrix}
0 & 1.02\times 10^{-3} & 1.02\times 10^{-3} & 1.02\times 10^{-3} \\
1.02\times 10^{-3} & 0 & 1.02\times 10^{-3} & 1.02\times 10^{-3} \\
1.02\times 10^{-3} & 1.02\times 10^{-3} & 0 & 1.02\times 10^{-3} \\
1.02\times 10^{-3} & 1.02\times 10^{-3} & 1.02\times 10^{-3} & 0 
\end{pmatrix}.
\end{equation*}
The matrix $C^l$ corresponds to a lower bound on the additional probability Bob has of causing a mismatch if he is dishonest, when compared to the honest case. For this reason we define the constant, non-zero element, as $guad$, the guaranteed advantage an honest strategy has over a dishonest strategy. Since $C^h_{ij}+C^l_{ij}\leq C_{ij}$, the minimum cost of $C^h+C^l$ must be less than the minimum cost of $C$. The matrix $C^l$ is of error-type, so its minimum cost is proportional to $p_{min}(\alpha)$, the minimum error probability, and can be achieved using a minimum-error measurement \cite{Wallden2013}. Since the optimal measurement is known, in this case, to be the SRM, we can calculate $p_{min}(\alpha)$ for all values of $\alpha$ from:

\begin{equation*}
\begin{split}
\lambda_{1} &= 2e^{- \left| \alpha\right|^2 } \left( \cos\left( \left| \alpha\right|^2 \right) + \cosh \left( \left| \alpha \right|^2 \right) \right)\\
\lambda_{2} &= 2e^{- \left| \alpha\right|^2 } \left( \sin\left( \left| \alpha\right|^2 \right) + \sinh \left( \left| \alpha \right|^2 \right) \right)\\
\lambda_{3} &= 2e^{- \left| \alpha\right|^2 } \left( \cosh\left( \left| \alpha\right|^2 \right) - \cos \left( \left| \alpha \right|^2 \right) \right)\\
\lambda_{4} &= 2e^{- \left| \alpha\right|^2 } \left( \sinh\left( \left| \alpha\right|^2 \right) - \sin \left( \left| \alpha \right|^2 \right) \right)\\
\end{split}
\end{equation*}

\begin{equation*}
p_{min}(\alpha) = 1-\frac{1}{16}\left| \sum\limits_{i} \sqrt{\lambda_{i}}\right|^{2}
\end{equation*}

Defining the gap, $g = p_{min}\times guad$, we know $p_h+g\leq C_{min}$. For the particular case of $|\alpha|^2 =0.4$, $p_{min} = 0.317$ and we get
\begin{equation*}
p_h + g = 8.61\times 10^{-5} + (0.317 \times 1.02 \times 10^{-3})
= 4.10\times 10^{-4} \leq C_{min}.
\end{equation*}
This is our lower bound on $C_{min}$ and in what follows we conservatively assume $C_{min} = p_h+g$. As long as we choose $s_v < C_{min}$, we can use \cite{HDF1963} to obtain
\begin{equation} \label{eq:for}
\mathbb{P}(\text{Forge}) \leq \exp(-(C_{min}-s_v)^2L).
\end{equation}

Note that for simplicity we have only considered the case of Bob attempting to forge. We should also consider the possibility of Charlie trying to forge. In this case, we would replace the cost matrix \eqref{eq:costmat} with the corresponding experimentally generated cost matrix for Charlie. The analysis then follows exactly as above and we would arrive at another value of $C_{min}$, valid for when Charlie is the forger. For the protocol to be secure against both Bob and Charlie attempting to forge, we would choose $C_{min} = \min\{C^{\text{Bob}}_{min}, C^{\text{Charlie}}_{min}\}$. For our implementation, $C^{\text{Bob}}_{min} < C^{\text{Charlie}}_{min}$ and so $C_{min}$ remains as above.

\subsection{Robustness}

Suppose all parties are honest. Bob aborts if either the $(1/2)L$ states received from Alice result in mismatch rate greater than $s_a$ or the measurement results for the $(1/2)L$ states received from Charlie give mismatch  rate greater than $s_a$. We suppose that the channel from Alice to Bob has an error rate of $p^B_h$ and the channel from Alice to Charlie has an error rate of $p^C_h$. Then we have 
$$ \mathbb{P}(\text{Abort due to Alice}) \leq \exp(-(s_a - p^B_h)^2L)$$
$$\mathbb{P}(\text{Abort due to Charlie}) \leq \exp(-(s_a - p^C_h)^2L). $$ 
If we set $p_h = \max(p^B_h, p^C_h)$ then the probability of an honest abort is bounded by
\begin{equation} \label{eq:honab}
\mathbb{P}(\text{honest abort}) \leq 2\exp(-(s_a-p_h)^2L).
\end{equation} 

\subsection{Signature Length}

Using the above analysis, we can calculate the length of the signature needed to securely sign a single-bit message. Following \cite{Collins2014}, we assume that there is no reason in general to favour one type of security over another, so we pick parameters $s_a$ and $s_v$ so as to make the probabilities of honest abort, forgery and repudiation all equal. If all these probabilities are all below $0.01\%$, we say that the protocol is secure to a level of $0.01\%$. By setting
\begin{equation}\label{eq:paramchoice}
s_a = p_h +g/4, ~~~ s_v = p_h + 3g/4,
\end{equation}
the probabilities of repudiation, forging, and honest abort all become approximately equal. More explicitly, considering the terms in the exponent of equations \eqref{eq:rep}, \eqref{eq:for}, \eqref{eq:honab}, we see they are equal when
\begin{equation} \label{eq:exponents}
\left(\frac{s_v-s_a}{2}\right)^2 - \frac{\ln(2)}{L} = (C_{min}-s_v)^2
= (s_a-p_h)^2 - \frac{\ln(2)}{L},
\end{equation}
where the pre-factors of 2 have been taken inside the exponential. As $L$ increases, the choice of $s_a$ and $s_v$ satisfying \eqref{eq:exponents} tends to those given in \eqref{eq:paramchoice}.

Determining the value of the coherent state amplitude, $\alpha$, that leads to the smallest possible signature length is, in general, a difficult problem. Higher values of $|\alpha|^2$ lead to lower loss and a smaller bit error rate, but come at the cost of making the states $\{ |\alpha\rangle, |i\alpha\rangle, |-\alpha\rangle, |-i\alpha\rangle \}$ more distinguishable for a potential forger. Thus we need to strike a balance between correctly transmitting the states, and giving power to a forger. In the previous QDS experiment \cite{Collins2014}, the magnitude of $\alpha$ was varied in increments of 1 between 1 and 10. In order to minimise the signature length, the magnitude of $\alpha$ is chosen so as to maximise the gap, $g$, defined above. It was found that $|\alpha|^2=1$ gave the largest gap for the range considered, and extrapolation of subsequent experimental data suggested that $|\alpha|^2 \approx 0.5$ would be optimal. In fact, we found in this experiment that the optimal value was around $|\alpha|^2=0.4$, for which the corresponding $p_{min}$ value is $0.317$.

We now have everything in place to calculate the signature length, $L$, needed to securely sign a message for $|\alpha|^2=0.4$. Over a distance of \SI{500}{\metre}, experimental data gives the honest cost as $p_h = \max(p^B_h, p^C_h)=1.26\times 10^{-4}$. Note that this differs from the value used in the analysis above because in fact Charlie's cost matrix gave a higher honest cost than Bob's. We also find $C_{min} = \min\{C^{\text{Bob}}_{min}, C^{\text{Charlie}}_{min}\} = 4.10 \times 10^{-4}$, as above. This gives the gap, $g=C_{min}-p_h=2.84\times 10^{-4}$, from which we can find the parameters $s_a, s_v$ using \eqref{eq:paramchoice}. Putting it all together, we find that a signature length of $L = 1.96\times 10^{9}$ is required to sign a message to a security level of $0.01\%$.

Although this is a significant improvement over the last QDS experiment \cite{Collins2014}, there are a number of ways to further improve the efficiency of the protocol. The simplest would be to increase the clock rate and therefore the transmission rate. This would not decrease the signature length, but would decrease the time needed to transmit a given signature length. The pulse rate used in this QDS system was \SI{100}{\mega\hertz}, and there is scope to increase this to $>$\SI{1}{\giga\hertz} as is typically found in modern QKD systems \cite{Sasaki2011, Dixon2010, Wang2012, Namekata2011}. Despite the clock rate increasing by a factor of 10, the increased effects of inter- and intra- symbol interference mean that the corresponding decrease in the signature time will not be as great as a a factor of 10.  The exact improvement strongly depends on the characteristics of the employed coherent source, single-photon detectors and timing electronics.

Another possible improvement would be to switch to an operational wavelength of \SI{1550}{\nano\metre} as in \cite{Dixon2010}, rather than the \SI{850}{\nano\metre} used in this experiment. At \SI{1550}{\nano\metre}, we would expect to see losses of about 0.22~dB per kilometer, a significant improvement over the 2.2~dB per kilometer in the current setup. This switch in wavelengths would enable a new QDS experiment to be carried out over $10$'s of kilometers with similar performance to our QDS system at short distances. A wavelength of \SI{850}{\nano\metre} was selected for these experiments as it provides a good compromise between the detection efficiency response of the mature low-noise, lowing timing-jitter, high efficiency thick junction silicon single-photon avalanche diodes at room temperature and the attenuation profile of fused silica optical fibres.

As the system used in these experiments is similar in layout to that used in a number of QKD experiments, we can make an approximate comparison of the performance of our QDS system with that of a state-of-the-art QKD system \cite{Dixon2010} by examining the quantum bit error rate (QBER). Reducing the QBER would decrease the size of the diagonal elements in the cost matrix \eqref{eq:costmat}. This would lead to a larger gap, $g$, and therefore a smaller signature. There is perhaps less scope for improvement in this respect since the current setup achieved a QBER of around $4\%$ over the range of distances investigated, which is comparable to the QKD system in \cite{Dixon2010} where the QBER at \SI{50}{\kilo\metre} was $3.85\%$.  The differing achievable transmission distances between the work presented here and reference \cite{Dixon2010} being due to the increased loss of the fused silica optical fibres for light with a wavelength of \SI{850}{\nano\metre}.

\section{Experimental Methods} \label{sec:method}

The experimental system shown in Figure \ref{Fig:experiment} of the main paper, shares many similarities with common phase-basis-set QKD experiments~\cite{GIS2002,Sasaki2011,Jouguet2013,Singh2014,Clarke2011b,Clarke2011}. The main components are similar to our previous \gls{USE} QDS system, with the multiport removed~\cite{Collins2014}. The sender Alice and receivers Bob and Charlie are all constructed from \SI{4.4}{\micro\meter} core diameter ``panda-eye'' polarization maintaining fiber~\cite{Nufern2013} that can support two orthogonal linear polarization modes. Alice generates \SI{850.17}{\nano\meter} wavelength coherent-state pulses at a repetition rate of \SI{100}{\mega\hertz} by means of an electrically gain-switched \gls{VCSEL} and a motorised optical attenuator~\cite{Collins2014}. A \gls{LiNbO3} phase modulator is used to encode the phases on weak pulses; a strong reference pulse is delayed by half a period from the encoded states. The uncertainty in the phase encoding is $\pm$\SI{1.6e-6}{\radian}, primarily induced by amplitude jitter in the electrical driving signal. The signal and reference pulses have orthogonal linear polarizations and are efficiently recombined using a \gls{PBC}~\cite{MAR1995} before transmission through the \SI{9}{\micro\meter} core diameter optical fiber quantum channel. The quantum channel is composed of reels of Corning SMF-28e optical fiber~\cite{CORSMF28e} that were retained within the same laboratory as Alice, Bob and Charlie. Short lengths of \SI{4.4}{\micro\meter} core diameter fiber~\cite{Nufern2013a} are spliced onto the input and output of the quantum channel to eliminate higher order spatial modes~\cite{GOR2004B}. 

Mechanical and thermal stresses induced on the quantum channel will result in a time-evolving birefringence in the quantum channel that serves to reduce the linearity of the polarizations as they propagate. Bob and Charlie employ paddle-type \gls{SPC} to apply an opposing birefringence over a short ($\approx$\SI{1}{\meter}) length of \SI{4.4}{\micro\meter} core diameter fiber, returning the states to linear prior to the measurement step. This correction is applied manually once before a set of measurements are recorded and is monitored during operation, with realignment when necessary. Future revisions of this system will employ automatic correction of polarization evolution by means of monitoring a fraction of the delayed bright reference pulse.

A polarization beam-splitter in each receiver ensures that the weak signal pulse traverses the long path with the \SI{5}{\nano\second} delay while the bright reference pulse traverses the short path. The polarization and intensity of the reference pulse are altered to match the signal pulse before the two pulses are recombined on a final 50:50 \gls{BC} where they interfere. \gls{USE}~\cite{Dunjko2014} is used to eliminate some possible phases for the signal states~\cite{Collins2014}. Commercially available free-running Geiger-mode (photon-triggering)~\cite{PER2005}  \glspl{Si-SPAD}~\cite{Buller2010} are employed as detectors, because they have reasonable detection efficiency ($\approx$40\%) for photons with a wavelength of \SI{850}{\nano\meter}, a low dark count rate ($\approx$\SI{300}{\hertz}) and a low afterpulsing probability ($\approx$0.5\%) when compared to semiconductor detectors used at the telecommunications wavelengths~\cite{Collins2010,Buller2014}. Although superconducting detectors have exhibited detection efficiencies of 93\%~\cite{Marsili2013} such devices typically must be cooled to temperatures around 4~K with cryogens or large closed cycle coolers \cite{Natarajan2012} during operation while semiconductor technologies usually operate at near room temperature with Peltier coolers ~\cite{Buller2010,Buller2014}.  There have been recent advances in semiconductor single-photon detector technologies for wavelengths around 1550~nm~\cite{Buller2014} and it is likely that future QDS experiments over Corning SMF-28e optical fiber~\cite{CORSMF28e} will employ these technologies to further enhance the transmission range. Detector trigger events are time-stamped with time interval resolution of \SI{1}{\pico\second} using commercially available electronics~\cite{Wahl2008}. Technical limitations of interfacing and computer control meant that the maximum event rate that could be recorded by the receivers was approximately \SI{4}{\mega\hertz}. The time-stamping electronics require a \SI{10}{\mega\hertz} reference signal that is provided by common rubidium clock that is shared between Alice and the receivers.

The air-gaps in each receiver consist of an immobile launching collimating lens and a collection lens attached to a linear piezo-electric-actuator with a total travel of $\approx$\SI{1.5}{\micro\meter} in steps of $\approx$\SI{15}{\nano\meter}. The receivers adjust the relative lengths of their measurement setup to ensure optimum interferometric visibility (typically 93\%). The receivers' demodulation systems had a mean attenuation of 6.96~dB.

Analysis of the raw time-stamp information is carried out by custom software written in MATLAB~\cite{Mathworks2014}. The time-stamps are filtered to discard those that occur outside of a window of $\pm$\SI{2}{\nano\second} centred on the expected arrival time of a pulse – a process that retains 80\% on average of the raw counts. The non-gated counts are from background noise, dark counts and non-interfering pulses. The custom software examines the detector firing patterns to perform \gls{USE}. The measurement records are then retained to allow for subsequent forwarding of measurement results from Bob to Charlie and vice versa. This step needs to be done in secret from Alice, and could thus be accomplished using a standard QKD link between Bob and Charlie; this was however not implemented in the current setup.



\bibliography{BibliographyFile}

\begin{thebibliography}{48}
\expandafter\ifx\csname natexlab\endcsname\relax\def\natexlab#1{#1}\fi
\expandafter\ifx\csname bibnamefont\endcsname\relax
  \def\bibnamefont#1{#1}\fi
\expandafter\ifx\csname bibfnamefont\endcsname\relax
  \def\bibfnamefont#1{#1}\fi
\expandafter\ifx\csname citenamefont\endcsname\relax
  \def\citenamefont#1{#1}\fi
\expandafter\ifx\csname url\endcsname\relax
  \def\url#1{\texttt{#1}}\fi
\expandafter\ifx\csname urlprefix\endcsname\relax\def\urlprefix{URL }\fi
\providecommand{\bibinfo}[2]{#2}
\providecommand{\eprint}[2][]{\url{#2}}

\bibitem[{\citenamefont{Stinson}(2006)}]{STI2006}
\bibinfo{author}{\bibfnamefont{D.~R.} \bibnamefont{Stinson}},
  \emph{\bibinfo{title}{{Cryptography: {Theory} and practice}}}
  (\bibinfo{publisher}{Chapman \& Hall/CRC}, \bibinfo{year}{2006}),
  \bibinfo{edition}{3rd} ed.

\bibitem[{\citenamefont{Rivest et~al.}(1978)\citenamefont{Rivest, Shamir, and
  Adleman}}]{RIV1978}
\bibinfo{author}{\bibfnamefont{R.~L.} \bibnamefont{Rivest}},
  \bibinfo{author}{\bibfnamefont{A.}~\bibnamefont{Shamir}}, \bibnamefont{and}
  \bibinfo{author}{\bibfnamefont{L.}~\bibnamefont{Adleman}},
  \bibinfo{journal}{Commun. ACM} \textbf{\bibinfo{volume}{21}},
  \bibinfo{pages}{120} (\bibinfo{year}{1978}).

\bibitem[{\citenamefont{Elgamal}(1985)}]{Elgamal1985}
\bibinfo{author}{\bibfnamefont{T.}~\bibnamefont{Elgamal}},
  \bibinfo{journal}{IEEE Trans. Inf. Theory} \textbf{\bibinfo{volume}{31}},
  \bibinfo{pages}{469} (\bibinfo{year}{1985}).

\bibitem[{\citenamefont{Johnson et~al.}(2001)\citenamefont{Johnson, Menezes,
  and Vanstone}}]{Johnson2001}
\bibinfo{author}{\bibfnamefont{D.}~\bibnamefont{Johnson}},
  \bibinfo{author}{\bibfnamefont{A.}~\bibnamefont{Menezes}}, \bibnamefont{and}
  \bibinfo{author}{\bibfnamefont{S.}~\bibnamefont{Vanstone}},
  \bibinfo{journal}{Int. J. Inf. Secur.} \textbf{\bibinfo{volume}{1}},
  \bibinfo{pages}{36} (\bibinfo{year}{2001}).

\bibitem[{\citenamefont{Shor}(1997)}]{SHO1997}
\bibinfo{author}{\bibfnamefont{P.~W.} \bibnamefont{Shor}},
  \bibinfo{journal}{SIAM J. Comput.} \textbf{\bibinfo{volume}{26}},
  \bibinfo{pages}{1484} (\bibinfo{year}{1997}).

\bibitem[{\citenamefont{Gottesman and Chuang}(2002)}]{Gottesman2002}
\bibinfo{author}{\bibfnamefont{D.}~\bibnamefont{Gottesman}} \bibnamefont{and}
  \bibinfo{author}{\bibfnamefont{I.}~\bibnamefont{Chuang}},
  \bibinfo{journal}{arXiv preprint quant-ph/0105032}  (\bibinfo{year}{2002}).

\bibitem[{\citenamefont{Andersson et~al.}(2006)\citenamefont{Andersson, Curty,
  and Jex}}]{Andersson2006}
\bibinfo{author}{\bibfnamefont{E.}~\bibnamefont{Andersson}},
  \bibinfo{author}{\bibfnamefont{M.}~\bibnamefont{Curty}}, \bibnamefont{and}
  \bibinfo{author}{\bibfnamefont{I.}~\bibnamefont{Jex}},
  \bibinfo{journal}{Phys. Rev. A} \textbf{\bibinfo{volume}{74}},
  \bibinfo{pages}{022304} (\bibinfo{year}{2006}).

\bibitem[{\citenamefont{Swanson and Stinson}(2011)}]{Swanson2011}
\bibinfo{author}{\bibfnamefont{C.}~\bibnamefont{Swanson}} \bibnamefont{and}
  \bibinfo{author}{\bibfnamefont{D.}~\bibnamefont{Stinson}}, in
  \emph{\bibinfo{booktitle}{Information Theoretic Security}}, edited by
  \bibinfo{editor}{\bibfnamefont{S.}~\bibnamefont{Fehr}}
  (\bibinfo{publisher}{Springer}, \bibinfo{address}{Berlin},
  \bibinfo{year}{2011}), chap.~\bibinfo{chapter}{10}, pp.
  \bibinfo{pages}{100--116}, \bibinfo{edition}{1st} ed.

\bibitem[{\citenamefont{Chaum and
  Roijakkers}(1991)}]{Chaum:1990:USD:646755.705359}
\bibinfo{author}{\bibfnamefont{D.}~\bibnamefont{Chaum}} \bibnamefont{and}
  \bibinfo{author}{\bibfnamefont{S.}~\bibnamefont{Roijakkers}}, in
  \emph{\bibinfo{booktitle}{Proceedings of the 10th Annual International
  Cryptology Conference on Advances in Cryptology}}, edited by
  \bibinfo{editor}{\bibfnamefont{A.}~\bibnamefont{Menezes}} \bibnamefont{and}
  \bibinfo{editor}{\bibfnamefont{S.~A.} \bibnamefont{Vanstone}}
  (\bibinfo{publisher}{Springer-Verlag}, \bibinfo{address}{London, UK},
  \bibinfo{year}{1991}), CRYPTO '90, pp. \bibinfo{pages}{206--214}.

\bibitem[{\citenamefont{Hanaoka et~al.}(2000)\citenamefont{Hanaoka, Shikata,
  Zheng, and Imai}}]{Hanaoka2000}
\bibinfo{author}{\bibfnamefont{G.}~\bibnamefont{Hanaoka}},
  \bibinfo{author}{\bibfnamefont{J.}~\bibnamefont{Shikata}},
  \bibinfo{author}{\bibfnamefont{Y.}~\bibnamefont{Zheng}}, \bibnamefont{and}
  \bibinfo{author}{\bibfnamefont{H.}~\bibnamefont{Imai}}, in
  \emph{\bibinfo{booktitle}{Advances in Cryptology — ASIACRYPT 2000}}, edited
  by \bibinfo{editor}{\bibfnamefont{T.}~\bibnamefont{Okamoto}}
  (\bibinfo{publisher}{Springer}, \bibinfo{address}{Kyoto, Japan},
  \bibinfo{year}{2000}), pp. \bibinfo{pages}{130--142}.

\bibitem[{\citenamefont{Wegman and Carter}(1981)}]{WEG1981}
\bibinfo{author}{\bibfnamefont{M.~N.} \bibnamefont{Wegman}} \bibnamefont{and}
  \bibinfo{author}{\bibfnamefont{J.~L.} \bibnamefont{Carter}},
  \bibinfo{journal}{J. Comput. Syst. Sci.} \textbf{\bibinfo{volume}{22}},
  \bibinfo{pages}{265} (\bibinfo{year}{1981}).

\bibitem[{\citenamefont{Clarke et~al.}(2012)\citenamefont{Clarke, Collins,
  Dunjko, Andersson, Jeffers, and Buller}}]{Clarke2012}
\bibinfo{author}{\bibfnamefont{P.~J.} \bibnamefont{Clarke}},
  \bibinfo{author}{\bibfnamefont{R.~J.} \bibnamefont{Collins}},
  \bibinfo{author}{\bibfnamefont{V.}~\bibnamefont{Dunjko}},
  \bibinfo{author}{\bibfnamefont{E.}~\bibnamefont{Andersson}},
  \bibinfo{author}{\bibfnamefont{J.}~\bibnamefont{Jeffers}}, \bibnamefont{and}
  \bibinfo{author}{\bibfnamefont{G.~S.} \bibnamefont{Buller}},
  \bibinfo{journal}{Nat. Commun.} \textbf{\bibinfo{volume}{3}},
  \bibinfo{pages}{1174} (\bibinfo{year}{2012}).

\bibitem[{\citenamefont{Collins et~al.}(2014)\citenamefont{Collins, Donaldson,
  Dunjko, Wallden, Clarke, Andersson, Jeffers, and Buller}}]{Collins2014}
\bibinfo{author}{\bibfnamefont{R.~J.} \bibnamefont{Collins}},
  \bibinfo{author}{\bibfnamefont{R.~J.} \bibnamefont{Donaldson}},
  \bibinfo{author}{\bibfnamefont{V.}~\bibnamefont{Dunjko}},
  \bibinfo{author}{\bibfnamefont{P.}~\bibnamefont{Wallden}},
  \bibinfo{author}{\bibfnamefont{P.~J.} \bibnamefont{Clarke}},
  \bibinfo{author}{\bibfnamefont{E.}~\bibnamefont{Andersson}},
  \bibinfo{author}{\bibfnamefont{J.}~\bibnamefont{Jeffers}}, \bibnamefont{and}
  \bibinfo{author}{\bibfnamefont{G.~S.} \bibnamefont{Buller}},
  \bibinfo{journal}{Phys. Rev. Lett.} \textbf{\bibinfo{volume}{113}},
  \bibinfo{pages}{040502} (\bibinfo{year}{2014}).

\bibitem[{\citenamefont{Bussi\`{e}res et~al.}(2013)\citenamefont{Bussi\`{e}res,
  Sangouard, Afzelius, de~Riedmatten, Simon, and Tittel}}]{Bussieres2013}
\bibinfo{author}{\bibfnamefont{F.}~\bibnamefont{Bussi\`{e}res}},
  \bibinfo{author}{\bibfnamefont{N.}~\bibnamefont{Sangouard}},
  \bibinfo{author}{\bibfnamefont{M.}~\bibnamefont{Afzelius}},
  \bibinfo{author}{\bibfnamefont{H.}~\bibnamefont{de~Riedmatten}},
  \bibinfo{author}{\bibfnamefont{C.}~\bibnamefont{Simon}}, \bibnamefont{and}
  \bibinfo{author}{\bibfnamefont{W.}~\bibnamefont{Tittel}},
  \bibinfo{journal}{J. Mod. Opt.} \textbf{\bibinfo{volume}{60}},
  \bibinfo{pages}{1519} (\bibinfo{year}{2013}), \eprint{1306.6904}.

\bibitem[{\citenamefont{Simon et~al.}(2010)\citenamefont{Simon, Afzelius,
  Appel, {Boyer De La Giroday}, Dewhurst, Gisin, Hu, Jelezko, Kr\"{o}ll,
  M\"{u}ller et~al.}}]{Simon2010}
\bibinfo{author}{\bibfnamefont{C.}~\bibnamefont{Simon}},
  \bibinfo{author}{\bibfnamefont{M.}~\bibnamefont{Afzelius}},
  \bibinfo{author}{\bibfnamefont{J.}~\bibnamefont{Appel}},
  \bibinfo{author}{\bibfnamefont{A.}~\bibnamefont{{Boyer De La Giroday}}},
  \bibinfo{author}{\bibfnamefont{S.~J.} \bibnamefont{Dewhurst}},
  \bibinfo{author}{\bibfnamefont{N.}~\bibnamefont{Gisin}},
  \bibinfo{author}{\bibfnamefont{C.~Y.} \bibnamefont{Hu}},
  \bibinfo{author}{\bibfnamefont{F.}~\bibnamefont{Jelezko}},
  \bibinfo{author}{\bibfnamefont{S.}~\bibnamefont{Kr\"{o}ll}},
  \bibinfo{author}{\bibfnamefont{J.~H.} \bibnamefont{M\"{u}ller}},
  \bibnamefont{et~al.}, \bibinfo{journal}{Eur. Phys. J. D}
  \textbf{\bibinfo{volume}{58}}, \bibinfo{pages}{1} (\bibinfo{year}{2010}),
  \eprint{1003.1107}.

\bibitem[{\citenamefont{Specht et~al.}(2011)\citenamefont{Specht, N\"{o}lleke,
  Reiserer, Uphoff, Figueroa, Ritter, and Rempe}}]{Specht2011}
\bibinfo{author}{\bibfnamefont{H.~P.} \bibnamefont{Specht}},
  \bibinfo{author}{\bibfnamefont{C.}~\bibnamefont{N\"{o}lleke}},
  \bibinfo{author}{\bibfnamefont{A.}~\bibnamefont{Reiserer}},
  \bibinfo{author}{\bibfnamefont{M.}~\bibnamefont{Uphoff}},
  \bibinfo{author}{\bibfnamefont{E.}~\bibnamefont{Figueroa}},
  \bibinfo{author}{\bibfnamefont{S.}~\bibnamefont{Ritter}}, \bibnamefont{and}
  \bibinfo{author}{\bibfnamefont{G.}~\bibnamefont{Rempe}},
  \bibinfo{journal}{Nature} \textbf{\bibinfo{volume}{473}},
  \bibinfo{pages}{190} (\bibinfo{year}{2011}).

\bibitem[{\citenamefont{Reim et~al.}(2011)\citenamefont{Reim, Michelberger,
  Lee, Nunn, Langford, and Walmsley}}]{Reim2011}
\bibinfo{author}{\bibfnamefont{K.~F.} \bibnamefont{Reim}},
  \bibinfo{author}{\bibfnamefont{P.}~\bibnamefont{Michelberger}},
  \bibinfo{author}{\bibfnamefont{K.~C.} \bibnamefont{Lee}},
  \bibinfo{author}{\bibfnamefont{J.}~\bibnamefont{Nunn}},
  \bibinfo{author}{\bibfnamefont{N.~K.} \bibnamefont{Langford}},
  \bibnamefont{and} \bibinfo{author}{\bibfnamefont{I.~A.}
  \bibnamefont{Walmsley}}, \bibinfo{journal}{Phys. Rev. Lett.}
  \textbf{\bibinfo{volume}{107}}, \bibinfo{pages}{053603}
  (\bibinfo{year}{2011}).

\bibitem[{\citenamefont{Dunjko et~al.}(2014)\citenamefont{Dunjko, Wallden, and
  Andersson}}]{Dunjko2014}
\bibinfo{author}{\bibfnamefont{V.}~\bibnamefont{Dunjko}},
  \bibinfo{author}{\bibfnamefont{P.}~\bibnamefont{Wallden}}, \bibnamefont{and}
  \bibinfo{author}{\bibfnamefont{E.}~\bibnamefont{Andersson}},
  \bibinfo{journal}{Phys. Rev. Lett.} \textbf{\bibinfo{volume}{112}},
  \bibinfo{pages}{040502} (\bibinfo{year}{2014}).

\bibitem[{\citenamefont{Wallden et~al.}(2013)\citenamefont{Wallden, Dunjko, and
  Andersson}}]{Wallden2013}
\bibinfo{author}{\bibfnamefont{P.}~\bibnamefont{Wallden}},
  \bibinfo{author}{\bibfnamefont{V.}~\bibnamefont{Dunjko}}, \bibnamefont{and}
  \bibinfo{author}{\bibfnamefont{E.}~\bibnamefont{Andersson}},
  \bibinfo{journal}{J. Phys. A} \textbf{\bibinfo{volume}{47}},
  \bibinfo{pages}{125303} (\bibinfo{year}{2013}).

\bibitem[{\citenamefont{Wallden et~al.}(2015)\citenamefont{Wallden, Dunjko,
  Kent, and Andersson}}]{Wallden2015}
\bibinfo{author}{\bibfnamefont{P.}~\bibnamefont{Wallden}},
  \bibinfo{author}{\bibfnamefont{V.}~\bibnamefont{Dunjko}},
  \bibinfo{author}{\bibfnamefont{A.}~\bibnamefont{Kent}}, \bibnamefont{and}
  \bibinfo{author}{\bibfnamefont{E.}~\bibnamefont{Andersson}},
  \bibinfo{journal}{Phys. Rev. A} \textbf{\bibinfo{volume}{89}},
  \bibinfo{pages}{042304} (\bibinfo{year}{2015}).

\bibitem[{\citenamefont{Gisin et~al.}(2002)\citenamefont{Gisin, Ribordy,
  Tittel, and Zbinden}}]{GIS2002}
\bibinfo{author}{\bibfnamefont{N.}~\bibnamefont{Gisin}},
  \bibinfo{author}{\bibfnamefont{G.}~\bibnamefont{Ribordy}},
  \bibinfo{author}{\bibfnamefont{W.}~\bibnamefont{Tittel}}, \bibnamefont{and}
  \bibinfo{author}{\bibfnamefont{H.}~\bibnamefont{Zbinden}},
  \bibinfo{journal}{Rev. Mod. Phys.} \textbf{\bibinfo{volume}{74}},
  \bibinfo{pages}{145} (\bibinfo{year}{2002}).

\bibitem[{\citenamefont{Sasaki et~al.}(2011)\citenamefont{Sasaki, Fujiwara,
  Ishizuka, Klaus, Wakui, Takeoka, Miki, Yamashita, Wang, Tanaka
  et~al.}}]{Sasaki2011}
\bibinfo{author}{\bibfnamefont{M.}~\bibnamefont{Sasaki}},
  \bibinfo{author}{\bibfnamefont{M.}~\bibnamefont{Fujiwara}},
  \bibinfo{author}{\bibfnamefont{H.}~\bibnamefont{Ishizuka}},
  \bibinfo{author}{\bibfnamefont{W.}~\bibnamefont{Klaus}},
  \bibinfo{author}{\bibfnamefont{K.}~\bibnamefont{Wakui}},
  \bibinfo{author}{\bibfnamefont{M.}~\bibnamefont{Takeoka}},
  \bibinfo{author}{\bibfnamefont{S.}~\bibnamefont{Miki}},
  \bibinfo{author}{\bibfnamefont{T.}~\bibnamefont{Yamashita}},
  \bibinfo{author}{\bibfnamefont{Z.}~\bibnamefont{Wang}},
  \bibinfo{author}{\bibfnamefont{A.}~\bibnamefont{Tanaka}},
  \bibnamefont{et~al.}, \bibinfo{journal}{Opt. Express}
  \textbf{\bibinfo{volume}{19}}, \bibinfo{pages}{10387} (\bibinfo{year}{2011}).

\bibitem[{\citenamefont{Jouguet et~al.}(2013)\citenamefont{Jouguet,
  Kunz-Jacques, Leverrier, Grangier, and Diamanti}}]{Jouguet2013}
\bibinfo{author}{\bibfnamefont{P.}~\bibnamefont{Jouguet}},
  \bibinfo{author}{\bibfnamefont{S.}~\bibnamefont{Kunz-Jacques}},
  \bibinfo{author}{\bibfnamefont{A.}~\bibnamefont{Leverrier}},
  \bibinfo{author}{\bibfnamefont{P.}~\bibnamefont{Grangier}}, \bibnamefont{and}
  \bibinfo{author}{\bibfnamefont{E.}~\bibnamefont{Diamanti}},
  \bibinfo{journal}{Nat. Photonics} \textbf{\bibinfo{volume}{7}},
  \bibinfo{pages}{378} (\bibinfo{year}{2013}).

\bibitem[{\citenamefont{Singh et~al.}(2014)\citenamefont{Singh, Gupta, and
  Singh}}]{Singh2014}
\bibinfo{author}{\bibfnamefont{H.}~\bibnamefont{Singh}},
  \bibinfo{author}{\bibfnamefont{D.}~\bibnamefont{Gupta}}, \bibnamefont{and}
  \bibinfo{author}{\bibfnamefont{A.}~\bibnamefont{Singh}},
  \bibinfo{journal}{IOSR J. Comput. Eng.} \textbf{\bibinfo{volume}{16}},
  \bibinfo{pages}{01} (\bibinfo{year}{2014}).

\bibitem[{\citenamefont{Clarke et~al.}(2011{\natexlab{a}})\citenamefont{Clarke,
  Collins, Hiskett, Townsend, and Buller}}]{Clarke2011b}
\bibinfo{author}{\bibfnamefont{P.~J.} \bibnamefont{Clarke}},
  \bibinfo{author}{\bibfnamefont{R.~J.} \bibnamefont{Collins}},
  \bibinfo{author}{\bibfnamefont{P.~A.} \bibnamefont{Hiskett}},
  \bibinfo{author}{\bibfnamefont{P.~D.} \bibnamefont{Townsend}},
  \bibnamefont{and} \bibinfo{author}{\bibfnamefont{G.~S.}
  \bibnamefont{Buller}}, \bibinfo{journal}{Appl. Phys. Lett.}
  \textbf{\bibinfo{volume}{98}}, \bibinfo{pages}{131103}
  (\bibinfo{year}{2011}{\natexlab{a}}).

\bibitem[{\citenamefont{Amiri et~al.}(2015)\citenamefont{Amiri, Wallden, Kent,
  and Andersson}}]{Amiri2015}
\bibinfo{author}{\bibfnamefont{R.}~\bibnamefont{Amiri}},
  \bibinfo{author}{\bibfnamefont{P.}~\bibnamefont{Wallden}},
  \bibinfo{author}{\bibfnamefont{A.}~\bibnamefont{Kent}}, \bibnamefont{and}
  \bibinfo{author}{\bibfnamefont{E.}~\bibnamefont{Andersson}},
  \bibinfo{journal}{arXiv preprint arXiv:1507.02975}  (\bibinfo{year}{2015}).

\bibitem[{\citenamefont{Carter and Wegman}(1979)}]{CAR1979}
\bibinfo{author}{\bibfnamefont{J.}~\bibnamefont{Carter}} \bibnamefont{and}
  \bibinfo{author}{\bibfnamefont{M.~N.} \bibnamefont{Wegman}},
  \bibinfo{journal}{J. Comput. Syst. Sci.} \textbf{\bibinfo{volume}{18}},
  \bibinfo{pages}{143} (\bibinfo{year}{1979}).

\bibitem[{\citenamefont{Abidin and Larsson}(2014)}]{Abidin2014}
\bibinfo{author}{\bibfnamefont{A.}~\bibnamefont{Abidin}} \bibnamefont{and}
  \bibinfo{author}{\bibfnamefont{J.-{\AA}.} \bibnamefont{Larsson}},
  \bibinfo{journal}{Quantum Inf. Process.} \textbf{\bibinfo{volume}{13}},
  \bibinfo{pages}{2155} (\bibinfo{year}{2014}).

\bibitem[{\citenamefont{Korzh et~al.}(2014)\citenamefont{Korzh, Lim, Houlmann,
  Gisin, Li, Nolan, Sanguinetti, Thew, and Zbinden}}]{Korzh2014}
\bibinfo{author}{\bibfnamefont{B.}~\bibnamefont{Korzh}},
  \bibinfo{author}{\bibfnamefont{C.~C.~W.} \bibnamefont{Lim}},
  \bibinfo{author}{\bibfnamefont{R.}~\bibnamefont{Houlmann}},
  \bibinfo{author}{\bibfnamefont{N.}~\bibnamefont{Gisin}},
  \bibinfo{author}{\bibfnamefont{M.~J.} \bibnamefont{Li}},
  \bibinfo{author}{\bibfnamefont{D.}~\bibnamefont{Nolan}},
  \bibinfo{author}{\bibfnamefont{B.}~\bibnamefont{Sanguinetti}},
  \bibinfo{author}{\bibfnamefont{R.}~\bibnamefont{Thew}}, \bibnamefont{and}
  \bibinfo{author}{\bibfnamefont{H.}~\bibnamefont{Zbinden}},
  \bibinfo{journal}{Nat. Photon.} \textbf{\bibinfo{volume}{9}},
  \bibinfo{pages}{7} (\bibinfo{year}{2014}), \eprint{1407.7427}.

\bibitem[{\citenamefont{Chv\'{a}tal}(1979)}]{Chvatal1979}
\bibinfo{author}{\bibfnamefont{V.}~\bibnamefont{Chv\'{a}tal}},
  \bibinfo{journal}{Discrete Math.} \textbf{\bibinfo{volume}{25}},
  \bibinfo{pages}{285} (\bibinfo{year}{1979}).

\bibitem[{\citenamefont{Hoeffding}(1963)}]{HDF1963}
\bibinfo{author}{\bibfnamefont{W.}~\bibnamefont{Hoeffding}},
  \bibinfo{journal}{J. Am. Stat. Assoc.} \textbf{\bibinfo{volume}{58}},
  \bibinfo{pages}{13} (\bibinfo{year}{1963}).

\bibitem[{\citenamefont{Dixon et~al.}(2010)\citenamefont{Dixon, Yuan, Dynes,
  Sharpe, and Shields}}]{Dixon2010}
\bibinfo{author}{\bibfnamefont{A.~R.} \bibnamefont{Dixon}},
  \bibinfo{author}{\bibfnamefont{Z.~L.} \bibnamefont{Yuan}},
  \bibinfo{author}{\bibfnamefont{J.~F.} \bibnamefont{Dynes}},
  \bibinfo{author}{\bibfnamefont{A.~W.} \bibnamefont{Sharpe}},
  \bibnamefont{and} \bibinfo{author}{\bibfnamefont{A.~J.}
  \bibnamefont{Shields}}, \bibinfo{journal}{Appl. Phys. Lett.}
  \textbf{\bibinfo{volume}{96}}, \bibinfo{pages}{2008} (\bibinfo{year}{2010}).

\bibitem[{\citenamefont{Wang et~al.}(2012)\citenamefont{Wang, Chen, Guo, Yin,
  Li, Zhou, Guo, and Han}}]{Wang2012}
\bibinfo{author}{\bibfnamefont{S.}~\bibnamefont{Wang}},
  \bibinfo{author}{\bibfnamefont{W.}~\bibnamefont{Chen}},
  \bibinfo{author}{\bibfnamefont{J.-F.} \bibnamefont{Guo}},
  \bibinfo{author}{\bibfnamefont{Z.-Q.} \bibnamefont{Yin}},
  \bibinfo{author}{\bibfnamefont{H.-W.} \bibnamefont{Li}},
  \bibinfo{author}{\bibfnamefont{Z.}~\bibnamefont{Zhou}},
  \bibinfo{author}{\bibfnamefont{G.-C.} \bibnamefont{Guo}}, \bibnamefont{and}
  \bibinfo{author}{\bibfnamefont{Z.-F.} \bibnamefont{Han}},
  \bibinfo{journal}{Opt. Lett.} \textbf{\bibinfo{volume}{37}},
  \bibinfo{pages}{1008} (\bibinfo{year}{2012}).

\bibitem[{\citenamefont{Namekata et~al.}(2011)\citenamefont{Namekata, Takesue,
  Honjo, Tokura, and Inoue}}]{Namekata2011}
\bibinfo{author}{\bibfnamefont{N.}~\bibnamefont{Namekata}},
  \bibinfo{author}{\bibfnamefont{H.}~\bibnamefont{Takesue}},
  \bibinfo{author}{\bibfnamefont{T.}~\bibnamefont{Honjo}},
  \bibinfo{author}{\bibfnamefont{Y.}~\bibnamefont{Tokura}}, \bibnamefont{and}
  \bibinfo{author}{\bibfnamefont{S.}~\bibnamefont{Inoue}},
  \bibinfo{journal}{Opt. Express} \textbf{\bibinfo{volume}{19}},
  \bibinfo{pages}{10632} (\bibinfo{year}{2011}).

\bibitem[{\citenamefont{Clarke et~al.}(2011{\natexlab{b}})\citenamefont{Clarke,
  Collins, Hiskett, Garc\'{\i}a-Mart\'{\i}nez, Krichel, McCarthy, Tanner,
  O'Connor, Natarajan, Miki et~al.}}]{Clarke2011}
\bibinfo{author}{\bibfnamefont{P.~J.} \bibnamefont{Clarke}},
  \bibinfo{author}{\bibfnamefont{R.~J.} \bibnamefont{Collins}},
  \bibinfo{author}{\bibfnamefont{P.~A.} \bibnamefont{Hiskett}},
  \bibinfo{author}{\bibfnamefont{M.-J.}
  \bibnamefont{Garc\'{\i}a-Mart\'{\i}nez}},
  \bibinfo{author}{\bibfnamefont{N.~J.} \bibnamefont{Krichel}},
  \bibinfo{author}{\bibfnamefont{A.}~\bibnamefont{McCarthy}},
  \bibinfo{author}{\bibfnamefont{M.~G.} \bibnamefont{Tanner}},
  \bibinfo{author}{\bibfnamefont{J.~A.} \bibnamefont{O'Connor}},
  \bibinfo{author}{\bibfnamefont{C.~M.} \bibnamefont{Natarajan}},
  \bibinfo{author}{\bibfnamefont{S.}~\bibnamefont{Miki}}, \bibnamefont{et~al.},
  \bibinfo{journal}{New J. Phys.} \textbf{\bibinfo{volume}{13}},
  \bibinfo{pages}{075008} (\bibinfo{year}{2011}{\natexlab{b}}).

\bibitem[{\citenamefont{Nufern}(2013{\natexlab{a}})}]{Nufern2013}
\bibinfo{author}{\bibnamefont{Nufern}}, \emph{\bibinfo{title}{{Polarization
  Maintaining Short Wavelength Fibers}}} (\bibinfo{year}{2013}{\natexlab{a}}),
  \urlprefix\url{http://www.nufern.com/pam/optical\_fibers/961/PM780-HP/}.

\bibitem[{\citenamefont{Marand and Townsend}(1995)}]{MAR1995}
\bibinfo{author}{\bibfnamefont{C.}~\bibnamefont{Marand}} \bibnamefont{and}
  \bibinfo{author}{\bibfnamefont{P.~D.} \bibnamefont{Townsend}},
  \bibinfo{journal}{Opt. Lett.} \textbf{\bibinfo{volume}{20}},
  \bibinfo{pages}{1695} (\bibinfo{year}{1995}).

\bibitem[{COR(2005)}]{CORSMF28e}
\emph{\bibinfo{title}{{Corning SMF-28e Optical Fiber Product Information}}}
  (\bibinfo{year}{2005}),
  \urlprefix\url{http://www.tlc.unipr.it/cucinotta/cfa/datasheet\_SMF28e.pdf}.

\bibitem[{\citenamefont{Nufern}(2013{\natexlab{b}})}]{Nufern2013a}
\bibinfo{author}{\bibnamefont{Nufern}}, \emph{\bibinfo{title}{{Nufern 780 nm
  Select Cut-Off Single-Mode fiber}}} (\bibinfo{year}{2013}{\natexlab{b}}),
  \urlprefix\url{http://www.nufern.com/pam/optical\_fibers/883/780-HP/}.

\bibitem[{\citenamefont{Gordon et~al.}(2004)\citenamefont{Gordon, Fernandez,
  Townsend, and Buller}}]{GOR2004B}
\bibinfo{author}{\bibfnamefont{K.~J.} \bibnamefont{Gordon}},
  \bibinfo{author}{\bibfnamefont{V.}~\bibnamefont{Fernandez}},
  \bibinfo{author}{\bibfnamefont{P.~D.} \bibnamefont{Townsend}},
  \bibnamefont{and} \bibinfo{author}{\bibfnamefont{G.~S.}
  \bibnamefont{Buller}}, \bibinfo{journal}{IEEE J. Quantum Electron.}
  \textbf{\bibinfo{volume}{40}}, \bibinfo{pages}{900} (\bibinfo{year}{2004}).

\bibitem[{\citenamefont{PerkinElmer}(2005)}]{PER2005}
\bibinfo{author}{\bibnamefont{PerkinElmer}}, \emph{\bibinfo{title}{{SPCM-AQR
  Single Photon Counting Module}}}, \bibinfo{howpublished}{Perkin Elmer
  Datasheet} (\bibinfo{year}{2005}),
  \urlprefix\url{www.optoelectronics.perkinelmer.com}.

\bibitem[{\citenamefont{Buller and Collins}(2010)}]{Buller2010}
\bibinfo{author}{\bibfnamefont{G.~S.} \bibnamefont{Buller}} \bibnamefont{and}
  \bibinfo{author}{\bibfnamefont{R.~J.} \bibnamefont{Collins}},
  \bibinfo{journal}{Meas. Sci. Technol.} \textbf{\bibinfo{volume}{21}},
  \bibinfo{pages}{012002} (\bibinfo{year}{2010}).

\bibitem[{\citenamefont{Collins et~al.}(2010)\citenamefont{Collins, Hadfield,
  and Buller}}]{Collins2010}
\bibinfo{author}{\bibfnamefont{R.~J.} \bibnamefont{Collins}},
  \bibinfo{author}{\bibfnamefont{R.~H.} \bibnamefont{Hadfield}},
  \bibnamefont{and} \bibinfo{author}{\bibfnamefont{G.~S.}
  \bibnamefont{Buller}}, \bibinfo{journal}{J. Nanophotonics}
  \textbf{\bibinfo{volume}{4}}, \bibinfo{pages}{040301} (\bibinfo{year}{2010}).

\bibitem[{\citenamefont{Buller and Collins}(2014)}]{Buller2014}
\bibinfo{author}{\bibfnamefont{G.~S.} \bibnamefont{Buller}} \bibnamefont{and}
  \bibinfo{author}{\bibfnamefont{R.~J.} \bibnamefont{Collins}}, in
  \emph{\bibinfo{booktitle}{Springer Series on Fluorescence: Methods and
  Applications: Advanced Photon Counting}}, edited by
  \bibinfo{editor}{\bibfnamefont{P.}~\bibnamefont{Kapusta}},
  \bibinfo{editor}{\bibfnamefont{M.}~\bibnamefont{Wahl}}, \bibnamefont{and}
  \bibinfo{editor}{\bibfnamefont{R.}~\bibnamefont{Erdmann}}
  (\bibinfo{publisher}{Springer}, \bibinfo{year}{2014}),
  chap.~\bibinfo{chapter}{3}.

\bibitem[{\citenamefont{Marsili et~al.}(2013)\citenamefont{Marsili, Verma,
  Stern, Harrington, Lita, Gerrits, Vayshenker, Baek, Shaw, Mirin
  et~al.}}]{Marsili2013}
\bibinfo{author}{\bibfnamefont{F.}~\bibnamefont{Marsili}},
  \bibinfo{author}{\bibfnamefont{V.~B.} \bibnamefont{Verma}},
  \bibinfo{author}{\bibfnamefont{J.~A.} \bibnamefont{Stern}},
  \bibinfo{author}{\bibfnamefont{S.}~\bibnamefont{Harrington}},
  \bibinfo{author}{\bibfnamefont{A.~E.} \bibnamefont{Lita}},
  \bibinfo{author}{\bibfnamefont{T.}~\bibnamefont{Gerrits}},
  \bibinfo{author}{\bibfnamefont{I.}~\bibnamefont{Vayshenker}},
  \bibinfo{author}{\bibfnamefont{B.}~\bibnamefont{Baek}},
  \bibinfo{author}{\bibfnamefont{M.~D.} \bibnamefont{Shaw}},
  \bibinfo{author}{\bibfnamefont{R.~P.} \bibnamefont{Mirin}},
  \bibnamefont{et~al.}, \bibinfo{journal}{Nat. Photon.}
  \textbf{\bibinfo{volume}{7}}, \bibinfo{pages}{210} (\bibinfo{year}{2013}),
  \eprint{1209.5774}.

\bibitem[{\citenamefont{Natarajan et~al.}(2012)\citenamefont{Natarajan, Tanner,
  and Hadfield}}]{Natarajan2012}
\bibinfo{author}{\bibfnamefont{C.~M.} \bibnamefont{Natarajan}},
  \bibinfo{author}{\bibfnamefont{M.~G.} \bibnamefont{Tanner}},
  \bibnamefont{and} \bibinfo{author}{\bibfnamefont{R.~H.}
  \bibnamefont{Hadfield}}, \bibinfo{journal}{Supercond. Sci. Technol.}
  \textbf{\bibinfo{volume}{25}}, \bibinfo{pages}{063001}
  (\bibinfo{year}{2012}), \eprint{1204.5560}.

\bibitem[{\citenamefont{Wahl et~al.}(2008)\citenamefont{Wahl, Rahn,
  R\"{o}hlicke, Kell, Nettels, Hillger, Schuler, and Erdmann}}]{Wahl2008}
\bibinfo{author}{\bibfnamefont{M.}~\bibnamefont{Wahl}},
  \bibinfo{author}{\bibfnamefont{H.-J.} \bibnamefont{Rahn}},
  \bibinfo{author}{\bibfnamefont{T.}~\bibnamefont{R\"{o}hlicke}},
  \bibinfo{author}{\bibfnamefont{G.}~\bibnamefont{Kell}},
  \bibinfo{author}{\bibfnamefont{D.}~\bibnamefont{Nettels}},
  \bibinfo{author}{\bibfnamefont{F.}~\bibnamefont{Hillger}},
  \bibinfo{author}{\bibfnamefont{B.}~\bibnamefont{Schuler}}, \bibnamefont{and}
  \bibinfo{author}{\bibfnamefont{R.}~\bibnamefont{Erdmann}},
  \bibinfo{journal}{Rev. Sci. Instrum.} \textbf{\bibinfo{volume}{79}},
  \bibinfo{pages}{123113} (\bibinfo{year}{2008}).

\bibitem[{\citenamefont{Mathworks}(2014)}]{Mathworks2014}
\bibinfo{author}{\bibnamefont{Mathworks}}, \emph{\bibinfo{title}{{MATLAB 2014b
  (8.4.0.118713)}}} (\bibinfo{year}{2014}),
  \urlprefix\url{http://mathworks.com/products/matlab/}.

\end{thebibliography}
\bibliographystyle{apsrev}


\end{document}